
\normalbaselineskip=12pt
\baselineskip=12pt
\hsize 15.0truecm \hoffset 1.0 truecm
\vsize 22.0truecm
\nopagenumbers
\headline={\ifnum \pageno=1
\hfil \else\hss\tenrm\folio\hss\fi}
\pageno=1

\def\lsim{\mathrel{\rlap{\lower4pt\hbox{\hskip1pt$\sim$}}
    \raise1pt\hbox{$<$}}}     
\def\gsim{\mathrel{\rlap{\lower4pt\hbox{\hskip1pt$\sim$}}
    \raise1pt\hbox{$>$}}}   

\hfill IUHET  270

\line{\hfil January, 1994}

\bigskip
\vskip 12pt
\centerline{\bf A New Method to Predict Meson Masses}
\vskip 36pt
\centerline{R. Roncaglia, A. R. Dzierba,
D. B. Lichtenberg, and
E. Predazzi,\footnote{$^{*}$}{On leave
from the University of Torino, Italy}}

\centerline{\it Department of Physics,
Indiana University, Bloomington, Indiana, 47405}

\vskip 1.0in

\item{}
The Feynman--Hellmann theorem is
used to show that vector meson energy eigenvalues
are monotonically decreasing
functions of the reduced masses of their constituent quarks.
The experimental meson masses are used
to put constraints on the values of quark masses and to
predict the masses of some as yet undiscovered mesons.
The mass of the $B_c^*$ meson is
predicted to be $6320\pm 10$ MeV, and, with less precision,
the masses of a number of excited vector mesons are also
predicted.

\bigskip
\vfill \eject

Some years ago, Quigg and Rosner [1] applied the
Feynman--Hellmann theorem [2,3]
to obtain information about the binding energy of
a quark-antiquark system. Specifically, they
showed that if the quark-antiquark
interaction does not depend on quark flavor,
the energy eigenvalues of the
Schr\"odinger equation are monotonically
decreasing functions of the reduced mass of the system.
Since then, several authors [4--6] have used this result
to obtain constraints on quark and hadron masses, usually
in the form of inequalities.
The Feynman--Hellman theorem has also been applied to some
relativistic wave equations [7] and used with other
assumptions to obtain inequalities among quark masses [8].
In the present work, we use
the Feynman--Hellmann theorem as
a new method to predict the masses of
some as yet undiscovered vector mesons.

Let us consider a Hamiltonian $H$, which depends on
a parameter $\lambda$. Then the Feynman--Hellmann
theorem states that
$$\partial E / \partial \lambda=
\langle \partial H /  \partial \lambda  \rangle,
\eqno(1)$$
where $E$ is an eigenvalue of $H$ and the expectation value
is taken with respect to the normalized eigenfunction
belonging to $E$.

The Quigg and Rosner result is obtained by
applying (1) to the nonrelativistic
Hamiltonian $H=p^2/(2\mu) +V$, where
$\mu$ is the reduced mass
and $V$ is a flavor-independent interaction.  Then
$$\partial E / \partial \mu=
-\langle p^2\rangle/ (2\mu^2)  <0, \eqno(2)$$
{\it i.e.},
$E$ decreases monotonically as $\mu$
increases because $p^2$ is a positive definite operator.
Of course, if $V$ depends on $\mu$ but
$\partial V/\partial\mu \leq 0$, then
$\partial E/\partial\mu < 0$ remains valid.

In the relativistic case, the two-body Hamiltonian
depends  on the quark and antiquark
masses $m_1$ and $m_2$ rather than just on $\mu$.
As an example [7], we consider the two-body
Salpeter Hamiltonian
$H(m_1,m_2)$, given by
$$H(m_1,m_2)=
\sum_i[({\bf p}_i^2+m_i^2)^{1/2}-m_i] +V(m_1,m_2),
\eqno(3)$$
where we have let the interaction $V$ depend explicitly
on the $m_i$ ($i=1,2$). Taking the partial derivative
with respect to $m_i$  and using (1), we obtain
$$\partial E / \partial m_i=
\langle m_i / (p_i^2+m_i^2)^{1/2}\rangle -1
+\langle \partial V /\partial m_i\rangle. \eqno(4)$$

We can see from Eq.\ (4) that if
$$\langle\partial V /\partial m_i \rangle \leq 0, \eqno(5)$$
then,
$$\partial E / \partial m_i<0. \eqno(6)$$

Although we have considered a specific
Hamiltonian with relativistic kinematics, it is
interesting to examine the consequences of
requiring that (6) be true provided (5) holds.
The next step is to note that if $\mu$ increases and
neither $m_1$ nor $m_2$ decreases, it follows from (6)
that
$$\partial E / \partial \mu<0. \eqno(7)$$
Thus, $E$ decreases when $\mu$ increases,
{\it provided neither of the quark masses decreases}.

Let us assume isospin invariance
and consider mesons with constituent quarks $q$ ($=u$
or $d$), $s$, $c$, and $b$ whose masses satisfy
the inequalities $m_q<m_s<m_c<m_b$. Then, from the
discussion above, we expect
$$E_{bb}<E_{cb}<E_{cc}<E_{sc}
<E_{ss}<E_{qs}<E_{qq}, \eqno(8)$$
where $E_{ij}$ is a particular eigenvalue for quark $i$
and antiquark $j$.  (The inequalities remain true if
we replace  $E_{cc}$ by $E_{sb}$,
$E_{sc}$ by $E_{qb}$, and $E_{ss}$ by $E_{qc}$.)
Thus, (7) holds for the mesons with eigenergies
as in (8), whereas this may not
be the case when one of the two quark masses increases
and the other decreases.

We now discuss when
(5) might be valid for quarkonium states
(for a review of quark potential models, see [9]).
The interaction $V$
can be written as $V_1+ V_2$, where $V_1$ is independent
of quark flavors and $V_2$ depends on flavor.
The term $V_1$ is the static quark-antiquark
potential, which is commonly assumed
to contain a Coulomb-like term, an
approximately linear confining term,
and  a constant term, all independent of flavor.

The term $V_2$ is much more uncertain than $V_1$. In
the Fermi--Breit approximation, $V_2$ contains
both  spin-dependent and spin-independent terms which
are explicitly functions of flavor through
the quark masses. However, most phenomenological treatments of
quarkonia have not needed the Fermi--Breit
spin-independent term [9], and we shall neglect it here.
In states with zero orbital angular momentum,
the most important spin-dependent term is the
colormagnetic interaction. In a relativistic
treatment [10], this term,
which we denote by $V_{cm}$, has the form
$$V_{cm}= f(r)\sigma_i\cdot \sigma_j/(m_im_j), \eqno(9)$$
where $\sigma_i$ are Pauli spin matrices and $f(r)$ is
a positive definite operator which in some approximation
is proportional to a smeared-out delta function.

If $V=V_1 + V_{cm}$, we obtain
$$ \langle \partial V/\partial m_i\rangle =-\langle f(r)
\rangle \langle\sigma_i\cdot \sigma_j\rangle/(m_i^2m_j).
\eqno(10)$$
As is well known,
$\langle\sigma_i \cdot \sigma_j\rangle$
is 1 for vectors and $-3$ for pseudoscalars. Then,
because $f(r)$ is a positive definite operator,
we see from (10)  that the vector mesons
satisfy (5) but the pseudoscalars do not.
We therefore expect the energy eigenvalues
of the vectors to be monotonically decreasing functions
of $\mu$. On the other hand, the pseudoscalars violate this
condition for small $m_i$, where the contribution
from (10) is large.

In addition to $V_{cm}$, other terms are expected to
contribute to $V_2$.  Among them are instantons
[11], which are apparently important primarily in states
with spin and orbital angular momentum zero
(pseudoscalar mesons). Instanton
contributions depend on flavor in a way which is
more subtle than just through quark masses. For
example, they are different in meson states with isospin
zero and one, even if they contain the same quarks.
Instantons also tend to mix the wave functions of
certain mesons, like the $\eta$ and $\eta'$
(which
contain both $q\bar q$ and $s \bar s$ in their wave
functions, and perhaps some glueball admixture as well).
Such states are unsuitable for
our scheme, as, in order to compute the reduced mass of
a system, we must know its quark content.
This is another reason why
in this paper we
focus on vector mesons.

We now turn to the experimental data [12--14] on
ground-state vector mesons to see how well
our expectations are borne out.
We cannot directly use the data on meson masses $M_{ij}$
to compute the energy eigenvalues $E_{ij}$ because the
latter are given by
$$E_{ij}=M_{ij}-m_i-m_j, \eqno(11)$$
and the quark masses $m_i$ are  {\it a priori}
considered unknown.
However, we can test our ideas with
a selection of constituent quark masses
which have appeared in the literature [4,10,15--22].  In
Table I we give various sets of quark masses (the list is
by no means complete). For comparison, we give
in the first row the
set of quark masses which we use in this work.

Using four different sets of input quark masses from
Table I,
we show in Fig.\ 1 how $E$ varies as a function
of $\mu$ for the ground-state vector mesons.
For completeness, we include mesons
containing  $q\bar b$, $q\bar c$, and $s\bar b$ as well as
$s\bar c$, $s\bar s$, and $c\bar c$ in order to see
how $E$ varies with $\mu$ when  $m_1$ increases and
$m_2$ decreases.
We include
the $\rho$, $K^*$, $\phi$, $D^*$, $D_s^*$, $B^*$,
$B_s^*$, $J/\psi$, and $\Upsilon$.
We omit the $\omega$ meson in Fig.\ 1,
because in our scheme it is degenerate with the $\rho$,
although it
is actually 15 MeV heavier. (Instead of choosing
the $\rho$, we could choose the $\omega$ or an average
of both without appreciably affecting our results.)
All the masses are taken from the Particle Data Group [12],
except the mass of the $B_s^*$, which comes from two
recent measurements [13,14] of the mass of the $B_s$ plus a
measurement of $m(B_s^*)-m(B_s)$ (which needs confirmation)
quoted in [12].

We see from Fig.\ 1 (a) and (b) that, with
{\it some} choices of quark masses, $E$
appears indeed to be
a monotonically decreasing function of $\mu$ for the
vector mesons. In Fig. 1 (c) and (d),
we see that with
{\it other} choices of quark
masses, $E$ does not behave
smoothly as a function of $\mu$.
Although the sets of quark masses used in (c) and (d)
seem {\it a priori} as reasonable as those of
sets (a) and (b), the Feynman-Hellmann theorem tells us that
they are poor choices.

We next discuss how we obtain constraints on the
quark masses and at the same time predict the
masses of as yet undiscovered
mesons. We assume that $E$ is a monotonically decreasing
function of $\mu$.
We also assume that $E(\mu)$ is smooth
enough to lie on a curve containing only a few parameters,
and we attempt to fit the data on ground-state vector
mesons with  simple three-parameter curves
(the quark masses are four additional parameters).
We vary these 7 parameters so as to obtain a best fit to
the data. We have tried
several different three-parameter curves in
making our fits to get some idea of the errors
involved in a particular choice. Specifically, we have used
exponential, displaced-hyperbolic, and
parabolic functions, and
found that we obtain comparable
fits to the data with all three. Furthermore, the
meson energies are quite stable to our choice of functions.

We show in Fig.\ 2 our fit to the vector
meson energies with an exponential curve.
These quark masses are based on a somewhat arbitrary
choice of 300 MeV for the mass of the $u$ and $d$ quarks.
We can get comparable fits to the data
for  quark masses which
differ from our choices by 100 MeV or more, but the mass
differences are much more constrained.

The fit of the vector meson energies
to the curve $E(\mu)$ allows us to
predict the $B_c^*$ mass. We find
$$m(B_c^*)=6320\pm 10 \quad {\rm MeV}, \eqno(12)$$
where we have estimated the theoretical error from
the spread in values obtained using
the above mentioned different functional forms for $E(\mu)$.

We next turn to excited meson states, confining ourselves
to vector mesons for the
reasons we have already discussed.
The data are considerably poorer for the excited
states. We use only the data shown in Table II, taken
from the Particle Data Group [12]. Some of the quantum
numbers of mesons in  Table II are not known
experimentally. We omit the excited
$\rho$ and $\omega$ states, because they differ
considerably in mass, and this may indicate appreciable
mixing.
Because the data for the excited vector meson
states are sparse, it is adequate to use a linear
fit to predict the masses of missing states, again using
quark masses in the first row of Table I.

We show in Table III our predictions for the
masses of vector meson excited states as well as the
mass of the $B_c^*$ ground state. The
errors are conservative estimates, based not only on
the errors in the data but also on the error incurred
by assuming a linear fit. In addition,
the fact that the eigenergy of the $D_s^*(2470)$, whose
quantum numbers have not been measured,
lies close to our linear fit lends support to
the assumption that this state is a vector.

In conclusion, although the
Feynman-Hellmann theorem gives us only an inequality
($\partial E/\partial \mu < 0$), we are able to use
it, in conjunction with the assumption that $E(\mu)$
behaves smoothly,
to make quantitative predictions about the masses
of as yet undiscovered mesons. In order to
make these predictions, we have not assumed any specific
functional form for the quark-antiquark interaction,
but only some mild and
general characteristics about its flavor dependence.

We should like to thank Boris Kopeliovich and Malcolm
Macfarlane for valuable discussions. This work was
supported in part by the U. S. Department of Energy
and in part by the U. S. National Science Foundation.
\vfill\eject

References
\bigskip

\item{[1]} C. Quigg and J.~L. Rosner, Phys. Rep. {\bf 56},
167 (1979).

\item{[2]}R.~P. Feynman, Phys. Rev. {\bf 56}, 340 (1939).

\item{[3]}H. Hellmann, Acta Physicochimica URSS
I, 6, 913 (1935); IV, 2, 225 (1936); Einf\"uhrung
in die Quantenchemie (F. Denticke,
Leipzig and Vienna, 1937) p. 286.

\item{[4]}R.~A. Bertlmann and S. Ono,
Phys. Lett. {\bf 96B}, 123 (1980).

\item{[5]} R.~A. Bertlmann and A. Martin,
Nucl. Phys. B {\bf 168}, 111 (1980),

\item{[6]} H. Grosse and A. Martin, Phys. Rep.
{\bf 60}, 341 (1980).

\item{[7]} D. B. Lichtenberg,  Phys.\ Rev.\ D {\bf 40},
4196 (1989).

\item{[8]} D. B. Lichtenberg,  Phys.\ Rev.\ D {\bf 40},
3675 (1989).

\item{[9]} D. B. Lichtenberg,  Int.\ J. Mod.\ Phys. A
{\bf 2}, 1669 (1987).

\item{[10]} S. Godfrey and N. Isgur, Phys.\ Rev.\
D {\bf 32},  189 (1985).

\item{[11]} W. H. Blask, U. Bohm, M. G. Huber,
B. Ch. Metzsch, and H. R.
Petry, Z. Phys.\ A {\bf 337},  327 (1990)
and references therein.

\item{[12]} Particle Data Group, K. Hikasa {\it et al.},
Phys.\ Rev.\ D {\bf 45}, S1 (1992).

\item{[13]} V. Scarpine for the CDF Collaboration,
{\it Advanced Study Conf. on Heavy Flavours},
Pavia, Italy (Sept 3--7,1993).

\item{[14]} ALEPH Collaboration, Phys.\ Lett.\ B
(to be published).

\item{[15]} E. Eichten et al., Phys.\ Rev.\ D {\bf 21},
203 (1980).

\item{[16]} D. Flamm and F. Sch\"oberl
{\it Introduction to the quark model of elementary
particles}, New York: Gordon and Breach (1982).

\item{[17]} B. Silvestre-Brac, Phys.\ Rev.\ D {\bf 46},
2179 (1992).

\item{[18]} L. P. Fulcher (unpublished). 

\item{[19]} R. K. Bhaduri, L. E. Cohler, and
Y. Nogami, Nuovo Cimento {\bf 65A}, 376 (1981).

\item{[20]} S. Ono, Phys.\ Rev.\ D {\bf 17}, 888 (1978).


\item{[21]} X. Song, Phys.\ Rev.\ D
{\bf 40}, 3655 (1989). 

\item{[22]} Yong Wang and D. B. Lichtenberg,
Phys.\  Rev.\ D {\bf 42}, 2404 (1990).


\vfill\eject

TABLE I. Values of quark constituent masses in MeV for
calculating meson energy eigenvalues from
experimental values of their masses. We show
in the first row the values of the quark masses
used in this work and, for comparison,
values used by some other authors in subsequent rows.

\vskip 12pt
$$\vbox {\halign {\hfil #\hfil &&\quad \hfil #\hfil \cr
\cr \noalign{\hrule}
\cr \noalign{\hrule}
\cr
Reference & $m_q$ & $m_s$ & $m_c$ & $m_b$\cr
\cr \noalign{\hrule}
\cr
This work & 300 & 440 & 1590 & 4920 \cr
[4] & 310 & 620 & 1910 & 5270 \cr
[10] & 220 & 419 & 1628 & 4977 \cr
[15] & 335 & 450 & 1840 &5170 \cr
[16]   & 350 & 500 & 1500 & 4700 \cr
[17]  & 330 & 550 & 1650 & 4715 \cr
[18] & 150 & 366 & 1320 & 4749 \cr
[19]  & 337 & 600 & 1870 & 5259 \cr
[20]  & 336 & 510 & 1680 & 5000 \cr
[21] & 270 & 600 & 1700 & 5000 \cr
[22] & 300 & 500 & 1800 & 5200 \cr
\cr \noalign{\hrule}
\cr \noalign{\hrule}
}}$$

\bigskip

TABLE II. Input masses of excited vector mesons
from the tables of the Particle Data Group [12].
We do not include the  $\psi(3770)$ because we
believe it to be a state with orbital angular momentum 2.

\vskip 12pt
$$\vbox {\halign {\hfil #\hfil &&\quad \hfil #\hfil \cr
\cr \noalign{\hrule}
\cr \noalign{\hrule}
\cr
Name &Quark content & $n~^{2S+1}L_J$ & Mass  (MeV) \cr
\cr \noalign{\hrule}
\cr
$K^*$ & $\bar qs$ & $2~^3 S_1$ & $1412\pm 12$ \cr
$\phi$ & $\bar ss$ & $2~^3S_1$  & $1680\pm 50$ \cr
$^{\dagger}D^*$ & $\bar qc$ & $2~^3S_1$  & $2469\pm 10$ \cr
$\psi$ & $\bar cc$ & $2~^3S_1$  & $3686\pm 1$ \cr
$\Upsilon$ & $\bar bb$ & $2~^3S_1$  & $10023\pm 1$ \cr
$K^*$ & $\bar qs$ & $3~^3 S_1$ & $1714\pm 20$ \cr
$\psi$ & $\bar cc$ & $3~^3S_1$  & $4040\pm 10$ \cr
$\Upsilon$ & $\bar bb$ & $3~^3S_1$  & $10355\pm 1$ \cr

\cr \noalign{\hrule}
\cr \noalign{\hrule}
}}$$
$^{\dagger}$Not in main meson table of the Particle Data
Group, and needs confirmation. The spin and parity of
this state have not been measured.
\vfill\eject

TABLE III. Predicted masses of as yet unobserved vector
mesons.

\vskip 12pt
$$\vbox {\halign {\hfil #\hfil &&\quad \hfil #\hfil \cr
\cr \noalign{\hrule}
\cr \noalign{\hrule}
\cr
Name &Quark content & $n~^{2S+1}L_J$ & Mass  (MeV) \cr
\cr \noalign{\hrule}
\cr
$B_c^*$ & $\bar bc$ & $1~^3 S_1$ & $6320\pm 10$ \cr
$B^*$   & $\bar qb$ & $2~^3 S_1$ & $5830\pm 30$ \cr
$D_s^*$ & $\bar sc$ & $2~^3S_1$  & $2620\pm 30$ \cr
$B_s^*$ & $\bar bs$ & $2~^3S_1$  & $5940\pm 30$ \cr
$B_c^*$ & $\bar bc$ & $2~^3 S_1$ & $6940\pm 30$ \cr
$\phi$ & $\bar s s$ & $3~^3 S_1$ & $1860\pm 30$ \cr
$D^*$ & $\bar qc$ & $3~^3S_1$  & $2860\pm 30$ \cr
$B^*$ & $\bar qb$ & $3~^3 S_1$ & $6190\pm 30$ \cr
$D_s^*$ & $\bar sc$ & $3~^3S_1$  & $2980\pm 30$ \cr
$B_s^*$ & $\bar sb$ & $3~^3S_1$  & $6300\pm 30$ \cr
$B_c^*$ & $\bar bc$ & $3~^3 S_1$ & $7290\pm 30$ \cr

\cr \noalign{\hrule}
\cr \noalign{\hrule}
}}$$
\bigskip
\noindent {\bf Figure captions}
\bigskip
FIG.\ 1. Energy eigenvalues of vector
mesons using four sets of quark masses from Table I:
(a) from Ref. [4], (b) from Ref. [10], (c) from Ref. [16], and
(d) from Ref. [17]. The experimental
values of the meson masses are from Refs.\ [12--14].
\medskip
FIG.\ 2. Energy eigenvalues of vector
mesons using our quark masses from the first row
of Table I and the
experimental masses from Refs.\ [12--14].
The open circles are the known vectors
and the solid circle is our prediction for the $B_c^*$.
The solid line is a fit
to the vector meson data with an exponential form,
resulting in $E(\mu)=754\ \exp(-\mu/1375)-506$, where
all constants are in MeV.

\bye